
\magnification=\magstep1   \openup1\jot  \vsize=9truein
\font\titlefont=cmcsc10 scaled \magstep2
\rightline {DAMTP R/94/34} \rightline {gr-qc/9409003}
\vskip .5truein
\centerline {\titlefont Nonlinearity in Quantum Theory}\centerline{
\titlefont and Closed Timelike Curves}  \vskip 1truein

\centerline {M.~J.~Cassidy} \vskip 1truein
\centerline
{\it Department of Applied Mathematics and Theoretical Physics}
\centerline {\it University of Cambridge}  \centerline {\it Silver
Street}  \centerline {\it Cambridge, CB3 9EW}  \centerline {\it England}

\vskip .25truein  \centerline {\it August 1994}  \vskip
1truein  \centerline {\bf Abstract}
 \bigskip
We examine consequences of the density matrix approach to quantum theory in the
context of a model spacetime containing closed timelike curves. We find that in
general, an initially pure state will evolve in a nonlinear way to a mixed
quantum
state. CPT invariance and the implications of this nonlinearity for the
statistical
interpretation of quantum theory are discussed.
 \vfill \eject

\baselineskip=20pt

\beginsection 1. Introduction

Recently there has been a lot of interest in the question of whether one can
formulate a
well defined quantum field theory on spacetimes containing closed timelike
curves (CTCs)
and if so, how the theory will behave. Studies of a class of spacetimes which
have well
behaved initial and final chronal regions and a compact region of CTCs in the
middle have yielded some interesting results. If an S matrix is defined
relating the
initial and final quantum states in the asymptotic regions, then it was found
[1] that
free
field scattering is unitary but for the case of interacting fields, the Feynman
propagator fails to
satisfy the identities (analogous to the flat space Cutkosky rules) which would
establish
perturbative unitarity of the S matrix. This paper is concerned with the
various
attempts
that have been made to try and recover a Copenhagen interpretation, consistent
with
observations made both before and after the region of CTCs, for interacting
fields
on these spacetimes.

In the context of generalised quantum mechanics, Hartle [2] suggests that a
consistent
probabilistic interpretation can be recovered by normalising the amplitudes
with a
factor
which depends on the initial state of the system. However, in doing so, he
introduces a
fundamental nonlinearity into quantum theory. The path integral approach will
be
reviewed briefly
in Section 2.

Deutsch [4] looked at the problem from the point of view of quantum computation
by
considering the world lines of a finite number of particles, whose quantum
states are
described by density matrices. In his approach, a density matrix corresponding
to a
pure initial state can evolve into a density matrix corresponding to a mixed
quantum
state. This shows that the evolution cannot be given by $SS^{\dagger}$, where
$S$ is a
unitary matrix. In Section 3, we compare the evolution of density matrices
(given by Deutsch's rules) in a
spacetime containing a compact region of CTCs with the standard density matrix
analysis in a black hole geometry. For the CTC spacetime, one might guess that
the
evolution would be described by a linear superscattering operator, \$, as in
the
black hole case but one of the results of this paper is to show that the
density matrix
evolves nonlinearly under the rules that Deutsch proposes. This nonlinearity is
a much
worse feature than loss of quantum coherence and calls into question the
statistical
interpretation of quantum theory.

In a recent paper, Politzer [5] applied Deutsch's general method to a specific
example,
which has received much attention due to its simplicity. In this spacetime, two
flat,
spacelike, 3 dimensional disks of radius $r$ are located in Minkowski space at
$t=0$ and
$t=T$, with the same spatial coordinates. To introduce CTCs, one simply
identifies the
upper surface of the bottom disk with the lower surface of the top disk. Also,
the lower
surface of the bottom disk is identified with the upper surface of the top
disk, in
order to
avoid free surfaces. In his analysis, Politzer considers the world lines of
only two
interacting fermions, one inside and one outside the radius $r$. This model
spacetime is
reviewed in Section 4.

In Section 5, we examine the evolution of density matrices in the
Deutsch/Politzer
example and find that, when the initial density matrix is pure, the evolution
is
nonlinear and hence superposition is lost. CPT invariance of the time machine
and the
implications of this nonlinearity for the evolution of initially mixed states
are
discussed in Section 6.

\beginsection 2. The Sum over Histories Approach

This section presents a review of Hartle's approach [2] to spacetimes
containing CTCs. In generalised quantum mechanics, developed for closed systems
by
Gell--Mann and Hartle [see, for example, 3], there is still a time ordering of
field
operators so in the case of spacetimes containing CTCs, Hartle tries to
circumvent this
problem by use of the sum over histories formalism.

As in Feynman's sum over histories prescription, one considers amplitudes for
particular 4 dimensional field configurations $\phi (x)$, given by:
$$\rm{Amplitude}
\propto e^{iS\lbrack\phi(x)\rbrack/ \hbar},$$ where $S$ is the action
functional. Sets
of these
configurations represent coarse grained histories and a decoherence functional
$D(\alpha;\alpha^{\prime})$ measures the amount of quantum interference between
two
histories
$\alpha,\alpha^{\prime} \in A$, where $A$ is an exhaustive set of histories for
the
closed
system.

If $D(\alpha;\alpha^{\prime})$ is negligibly small for all pairs of different
$\alpha,\alpha^{\prime}\in A$, then the set is said to decohere and
probabilities can be
assigned to the individual histories, given by $p(\alpha) = D(\alpha;\alpha)$.
However, if $D(\alpha;\alpha^{\prime})\ne 0$ for any $\alpha,\alpha^{\prime}\in
A$, then
that means there is nonnegligible quantum interference between $\alpha$ and
$\alpha^{\prime}$, so there is no probabilistic interpretation for that
particular set
of
coarse grained histories.

Familiar Hamiltonian quantum mechanics can be cast in this form. The
dynamics is described by a characteristic decoherence functional,
$D(\alpha;\alpha^{\prime})$ and all the
fundamental principles of quantum mechanics ({\it e.g.} superposition,
hermiticity,
positivity) can be related to certain properties of
$D(\alpha;\alpha^{\prime})$. Other
decoherence functionals are acceptable in this framework as long as these basic
properties are respected. Therefore, by suitably changing the form of
$D(\alpha;\alpha^{\prime})$, one can generalise Hamiltonian quantum mechanics.

This is exactly what Hartle does in a recent paper [2] to try and recover a
reasonable
probabilistic interpretation of quantum theory in the presence of nonunitary
evolution. He considers a single scalar field in a general spacetime containing
a
compact region of CTCs. The chronal regions can still be foliated with
spacelike
hypersurfaces and on each surface, a Hilbert space of states can be defined.
Transition
 amplitudes between field configurations on
spacelike surfaces before and after the nonchronal region are given by a
typical path
integral over all 4--dimensional field configurations which match the initial
and final
conditions. Therefore, an S matrix defining scattering through the region of
CTCs can
be defined and calculated. However, the fact that the spacetime is not globally
hyperbolic implies that the S matrix is nonunitary (the propagator no longer
has the
form of half advanced minus retarded Green functions and hence does not satisfy
the
unitarity identities [1]).

Hartle's proposal for a consistent probabilistic interpretation is to introduce
a
decoherence functional which reduces to that of ordinary Hamiltonian quantum
mechanics
when there are no CTCs, but provides a generalised theory otherwise. This is
given
by: $$D(\beta^{\prime},\alpha^{\prime};\beta,\alpha) = {
{\rm tr}(C_{\beta^{\prime}}XC_{\alpha^{\prime}} \rho C_{\alpha}^{\dagger}
X^{\dagger}
C_{\beta}^
{\dagger}) \over {\rm tr}(X \rho X^{\dagger})}\space.$$

In the above equation, X is an operator describing the nonunitary evolution
through
the nonchronal region and $\rho$ is the
density matrix for the initial state of the system.

The $C_{\alpha},_{\beta}$ represent chains of unitary evolution operators and
projection operators in the chronal regions before/after the region of CTCs.
The
initial state evolves unitarily between the spacelike surfaces and the
projections are
on to relevant
subspaces of the Hilbert spaces defined on these surfaces. This functional does
lead
to consistent assignment of probabilities but unfortunately
the normalisation factor in the denominator renders the theory nonlinear in the
initial density matrix $\rho$.

If one takes a quantum cosmological point of view, where the unique initial
state
might be given by the Hartle-Hawking no boundary state, for example, then this
is not
such a problem but even so it is still an undesirable feature of the theory.
For
example, if an advanced civilisation were to create a wormhole sometime in the
future,
one can see (in principle at least) that evolved information from nonlinear
effects detected now might prevent the civilisation from ever building the
wormhole in
the first place!

Anderson [6] also proposed a rule for quantum evolution in the presence of
CTCs. The
essence of this proposal involves a `renormalisation' of the measure density
for the
Hilbert space of states defined after the acausal region. This simple idea
provides a
quantum mechanics which is causal (in the sense that probabilities in the
present are
not affected by achronal regions of spacetime in the future) and also linear in
the
initial density matrix.

However, Fewster and Wells [7] found a problem with this procedure on a
physical level.
In normal quantum mechanics, we think of a physical observable as an operator
which is
self-adjoint with respect to the inner product of the Hilbert space. In the
context of
Anderson's proposal, one would expect that observables must be represented by
operators which are self-adjoint with respect to both inner products. However,
it is
shown that as a result of this reqirement, allowed operators must commute with
the
nonunitary part
of the evolution, and their expectation values will only evolve according to
the
unitary part. This is a severe restriction:- for example, one can argue that
momentum
is unlikely to be in the class of allowed observables in this proposal.

Fewster and Wells suggest an alternative way to restore unitarity based on the
theory of
unitary dilations. They replace the nonunitary evolution operator $X$ by a
unitary
 dilation of $X$, which maps between enlarged (and possibly indefinite)
inner product spaces. In essence, the extra dimensions which are introduced
provide a
 space of states
for the particles in the CTC region which is inaccessible to outside observers
and so
 overall, global unitarity is maintained. However, when observers reduce the
state on
the
enlarged Hilbert space to the part they can measure, there will be loss of
quantum
coherence.

\beginsection 3. Mappings between tensor product Hilbert spaces

Consider a quantum system composed of two subsystems labelled 1 and 2. If there
are no
interactions between the subsystems, the total Hilbert space of states can be
written
as a tensor product: $$\cal{H} = \cal{H}_{\rm 1} \otimes \cal{H}_{\rm 2}~~~~.$$
In this
paper, we shall be interested in mappings between Hilbert spaces of this form.
If an
observer can only measure one part of the system, then all possibilities for
the
second subsystem must be summed over and in general, the resulting state
relevant for
the observer will be mixed.

The natural context in which to describe mixed states is given by the density
matrix
formalism. A density matrix $\rho$ describing a quantum system can be written
as:
$$\rho = \rho^A{}_B \vert A \rangle\langle B \vert~~~~ ,$$ where summation over
A and
B is
implied, $\lbrace\vert A \rangle\rbrace$ is a basis for $\cal H$ and
$\lbrace\langle B
\vert\rbrace$ is a basis for the complex conjugate Hilbert space
$\overline{\cal H}$.
The $\rho^A{}_B$ can be thought of as the components of a $(^1_1)$
tensor on Hilbert space or alternatively as a positive semi-definite matrix
with unit
trace. In a basis
in which $\rho^A{}_B$ is diagonal, we can write:$$\rho^A{}_B = \delta^A{}_B P_A
{}~~\rm{(no~ implicit~summation)}~~~~,$$ where $P_A$ is the probability for the
system
to be in the
 pure state $\vert A
\rangle$. If any $P_A = 1$ with all the others zero, then the system is said to
be in
a pure state. Otherwise, $\rho$ describes a mixed state ({\it i.e.} a sum of
pure states
weighted with probabilities $P_A$ such that $\sum_A P_A = 1$).

In the usual situation where a system is described by density matrices $\rho_-$
and
$\rho_+$ in asymptotic past and future regions, the superscattering operator
$\$:\rho_- \rightarrow \rho_+$ provides a unique linear mapping$$\rho_+{}^A{}_B
=
\$^A{}_{BC}{}^D~\rho_-{}^C{}_D$$between density matrices. $\$^A{}_{BC}{}^D$ is
Hermitian in each of its pairs of indices (AB,CD), probability is
conserved, {\it i.e.}$$\$^A{}_{AC}{}^D = \delta_C{}^D~~~~,$$ and when there is
no loss
of
quantum coherence ({\it i.e.} a pure state evolves to a pure state), it can be
factorised into
a product of a standard unitary $S$ operator and its adjoint,
$$\$^A{}_{BC}{}^D =
S^A{}_C \overline{S}_B{}^D~~~~.$$One can see how, in this approach,
nonunitarity of
the $S$ matrix is reinterpreted by saying that the system loses quantum
coherence in
the evolution and transition probabilities are given by the superscattering
operator,
\$, instead of
$\vert S\vert^2$. A more detailed analysis of \$ in the context of the
Euclidean approach to quantum gravity is given in [8]. We note here that one
can really only use \$ if the evolution of density matrices is linear. In this
paper, nonlinear evolution operators will be denoted by
$\Omega:\rho_-\rightarrow\rho_+$.

Locally, density matrices evolve according to the Schr\"odinger
equation:$$U\rho_-U^{\dagger}=\rho_+~~~~,$$where $U$ is a standard unitary
evolution
operator. We can write this equation in a component notation which will be
useful for
discussing density matrices on tensor product Hilbert
spaces:$$U^{~EF}_{~AB}\left(\rho_-\right)^{AB}_{CD}U^{\dagger ~CD}_{~~GH} =
\left(\rho_+\right)^{EF}_{GH}.$$ If $\rho_-$ is a density operator on ${\cal
H}^-$, and
the system can be decomposed as ${\cal H}^- = {\cal H}^-_1 \otimes {\cal
H}^-_2$, then
we
can write:$$\rho_- =  \rho_1^-\otimes\rho_2^-~~~~{\rm or}$$
$$(\rho_-)^{AB}_{CD} = (\rho_1^-)^A_C
(\rho_2^-)^B_D.$$ Even if an operator cannot be decomposed in this form
(like, for example, the operator U above), it
is still useful to think of the first (second) index in a pair as referring to
subsystem
1 (2). A pair of indices can be thought of as a single index referring to the
system
as a
whole.

As an example, let us consider the case of a black hole spacetime. In this
example, we
can write both the initial and final Hilbert spaces as tensor products:$${\cal
H}^- =
{\cal H}_1^-\otimes {\cal H}_2^-~~{\rm and}$$ $${\cal H}^
+ = {\cal H}_1^+ \otimes {\cal H}_2^+.$$

After the formation of the black hole, ${\cal H}_2^+$ is interpreted as the
Hilbert
space
of states for the particles hidden behind the event horizon. Before the black
hole
forms,
we can still define ${\cal H}_2^-$, but the density operator on this space is
assumed to
be in the vacuum state, ie $\rho_2^- = \vert O_2^-\rangle\langle O_2^-\vert$.
The total
initial density matrix will be given by $$\rho_- = \vert A_1^-\rangle\vert
O_2^-\rangle\langle B_1^-\vert\langle O_2^-\vert,$$ and in component form, the
total
evolution equation is $$\left(\rho_+\right)^{CD}_{EF} =
U^{~CD}_{~AO}\left(\rho_-\right)^{AO}_{BO}U^{\dagger ~BO}_{~~EF}.$$

The particle states behind the event horizon are unobservable, so we must sum
over all
possibilities to obtain the reduced superscattering operator relevant for
subsystem 1,
the region outside the horizon:$$\$^C{}_{EA}{}^B =
U^{~CD~}_{~AO~}U^{\dagger ~BO}_{~~ED}.$$

To emphasize the point made earlier, the mapping provided by this
$\$^C{}_{EA}{}^B$ is
only defined for subsystem 1 and the final reduced density operator obtained,
$\rho_1^+$, will in general correspond to a mixed state, even though the full
density
operator describes a pure state on ${\cal H}^+$.

We now apply this formalism to the evolution of density matrices on a spacetime
containing a compact region of CTCs. Deutsch's proposed rules for the evolution
consist of:

(a) an assumption that if the initial quantum state outside the time
machine, $\rho_1^-$, is pure at $t=0$, then the total quantum state can be
written as
a tensor product $\rho_-=\rho_1^-\otimes\rho_2^-$. The final quantum state at
$t=T$ is
not, in general, a tensor product but we can still obtain the density matrices
relevant for the individual subsystems by tracing over the degrees of freedom
inside/outside the time machine,

(b) an imposed boundary condition to obtain consistency around the
CTC:$$\left(\rho_+\right)^{EF}_{EH}\equiv (\rho_2^+)^F_H =
U^{~EF}_{~AB}\left(\rho_-\right)^{AB}_{CD}U^{\dagger~CD}_{~~EH} =
(\rho_2^-)^F_H~.$$

We wish to obtain an expression for the operator which maps initial to
final density matrices, just as in the black hole case. Generically, the
final external state $\rho_1^+$ will correspond to a mixed state and so there
is a
loss of quantum coherence as in the black hole case. However, the imposed
consistency condition creates an essential difference. When this boundary
condition is
 solved, one finds that the components $(\rho_2^-)^A_B$ are, in general, a
nonlinear
function of $(\rho_1^-)^C_D$ so this implies
that the evolution operator obtained on the basis of the above rules is
nonlinear.

The final state outside the region of CTCs is given
by:$$\left(\rho_1^+\right)^E_G = U^{~EF}_{~AB}\left(\rho_-\right)^{AB}_{CD}
U^{\dagger~CD}_{~~GF}$$
$$~~~~~~~~~~~~~~~~~~=U^{~EF}_{~AB}(\rho_1^-)^A_C(\rho_2^-)^B_D
U^{\dagger ~CD}_{~~GF}.$$ Hence, the reduced evolution operator relevant for
the
state outside the time machine is $$ \Omega^E{}_{GA}{}^C =
U^{~EF}_{~AB}(\rho_2^-)^B_D U^{\dagger ~CD}_{~~GF}.$$By inspection, one can see
that
$\Omega$ depends on $(\rho_2^-)^A_B$ which
implies that the overall scattering will also be nonlinear.

So far, the discussion has been fairly general. In a recent paper [5], Politzer
applies Deutsch's density matrix proposal to a specific spacetime containing
CTCs. In
his example, pure initial states generically evolve to
mixed states. The evolution of initially mixed states is treated by employing
the
standard statistical interpretation, {\it i.e.} by decomposing the
state into its constituent weighted pure states, evolving each pure state
separately
and then recombining them in the appropriate way at the end. In Section 5, we
focus
on a specific example of the nonlinear evolution discussed above in the context
of
Politzer's spacetime and consider
the implications for the evolution of initially mixed states in Section 6.
First of
all, though, we shall briefly review the Deutsch method applied to Politzer's
model
spacetime.

\beginsection 4. The Deutsch/Politzer example

 The example consists of a
fermion with fixed spatial coordinate and world line defined for all time $t$,
labelled subsystem 1. For $t$ between 0 and capital $T$, where $T>0$, the
fermion can
interact with an identical particle at a different fixed spatial location,
subsystem 2. For simplicity, we have assumed that system 2 does not exist for
$t > T$
or $t < 0$.

 In this region of interaction, the Hilbert space for the system is
4-dimensional, a suitable basis of states is given by the
$\lbrace\vert\uparrow\uparrow\rangle, \vert\uparrow\downarrow\rangle,
\vert\downarrow\uparrow\rangle, \vert\downarrow\downarrow\rangle\rbrace$ basis
and the state of the whole system at time $t$ can be described by a 4x4 density
matrix $\rho (t)$. The great advantage of considering fermions in this
example is that all calculations are reduced to 4x4 matrix manipulations. So
for example, by taking partial traces, we can obtain the 2x2 matrices
describing
the state of each individual particle. For the external particle, this is given
by $\rho_1(t)$ or ${\rm tr}_2 \rho(t)$, and
similarly the CTC particle is described by $\rho_2 (t)$.

Certain matching conditions need to be imposed. The boundary condition on
$\rho_2$
that brings the
CTC into the system is given by$$\rho_2(T^-) = \rho_2(0^+)$$ where $0^+ \equiv
0 + \epsilon$ and $T^- \equiv T - \epsilon$ (where $\epsilon$ is small). This
is just
rule (b) of Section 3 which enforces
consistency of information around the CTC. The other matching conditions are
given
by$$\rho_1(0^-) = \rho_1(0^+)~~~~~~\rm{,}$$ $$\rho_2(0^-) =
\rho_2(T^+)~~\rm{and}$$
$$\rho_1(T^+) = \rho_1(T^-)~~~~~~\rm{.}$$

Evolution of the density matrices is given by the Schr\"odinger equation
$$i\dot \rho = \lbrack H,\rho\rbrack\quad\hbox{or alternatively}\quad \rho (t)
=
U(t)\rho (0)U(t)^{-1}.$$
The evolution matrix $U(t) = e^{-iHt}$ is unitary, the evolution is linear
in $\rho$, and in the above basis, the Hamiltonian can be written in matrix
form
as:$$H = \left(\matrix{2\omega_0 +
\lambda&\gamma_1&\gamma_2&0\cr\gamma_1&\omega_0&\omega_1&0\cr\gamma_2&\omega_1&\omega_0
&0\cr0&0&0&0\cr}
\right),$$ where $\omega_0$ is like a mass parameter, $\omega_1$ corresponds to
a kinetic parameter, $\gamma_1$ and $\gamma_2$ represent Yukawa 3--particle
coupling
parameters, and $\lambda$ corresponds to a 4--fermion coupling parameter.

\beginsection 5. Evolution of the Density Matrices

We now look at what happens when the external particle evolves
towards $t=0$ and beyond in a pure quantum state, according to Deutsch's
prescription.
So we wish to use the boundary condition and the Schr\"odinger evolution to
determine $\rho_1(T^+)$ for a given $\rho_1(0^-)$. According to rule (a) of
Section 3,
if $\rho_1(0^-)$ is pure (in other words it can be written as $\rho_1(0^-) =
\vert\psi\rangle\langle\psi\vert$ where $\vert\psi\rangle$ is some quantum
state),
then at $t=0$ the most general 4x4
density matrix describing the system can be written in this tensor product
form:
$$\rho(0^-) = \rho_1(0^-)\otimes \left(\matrix{a&b\cr
b^{\ast}&(1-a)\cr}\right)~~~~.$$ The 2x2 matrix is just $\rho_2(0^+)$ and $a$
and
$b$ are undetermined parameters.

The final output will be given by $$\rho_1(T^+) =
{\rm tr}_2~\left[~ U~ \left[~\rho_1(0^-)\otimes \left(\matrix{a&b\cr
b^{\ast}&(1-a)\cr}\right)\right]~ U^{\dag}~\right]$$ and the
parameters $a$ and $b$ are determined by the constraint equations which are
just
another way of writing the boundary condition: $${\rm tr}_1~\left[~ U~
\left[~\rho_1(0^-)\otimes \left(\matrix{a&b\cr
b^{\ast}&(1-a)\cr}\right)\right]~
U^{\dag}~\right] = \left(\matrix{a&b\cr b^{\ast}&(1-a)\cr}\right).$$ In this
case when
the initial state is pure, the boundary condition provides enough constraints
to
determine $\rho_1(T^+)$ uniquely.

As Deutsch noted, the boundary condition can be written as $$\$\rho_2(0^+) =
\rho_2(0^+)~~~~,\rm{where}$$ $$\$ * = {\rm tr}_1 \left[U\left[\rho_1(0^-)
\otimes *
\right]
U^{\dagger} \right]$$ (* denotes the position of the operand).

In other words, $\rho_2$ is a fixed point of \$, where \$ is given by the above
and is
a linear superscattering operator on the space of density matrices describing
subsystem 2. In his paper, he proves that every operator of this form has a
fixed
point so there will always be a solution to this boundary condition and hence
in his
words - ``...closed timelike lines place no retrospective constraint on the
state of a
quantum system.''

We now give a specific example to show that the map from initial to final
density
matrices (for subsystem 1) is in fact nonlinear as described in Section 3.
We assume the tensor product form:$$\rho(0^+) = \rho_1(0^+) \otimes
\rho_2(0^+)~~~~,$$ where $\rho_1$ describes the initial (pure) quantum state of
the
external particle
and $\rho_2$ describes the CTC particle. The final 4x4 density matrix, and
hence
$\rho_1(T^+)$, are determined by both $\rho_1$ and $\rho_2$. However, as was
discussed
in Section 3, $\rho_2(0^+)$ depends nonlinearly on $\rho_1(0^+)$ and hence the
output
$\rho_1(T^+)$ will also be a nonlinear function of the initial state.

For example, suppose we write $$\rho(0^+) =
\left(\matrix{\alpha&\beta\cr\beta^*&\gamma\cr}\right) \otimes
\left(\matrix{a&b\cr b^*&c\cr}\right)~~~~,$$ where $\gamma = 1 - \alpha$,
$c=1-a$,
$\alpha$ and $\beta$ are fixed parameters describing the initial state and $a$
and $b$
are initially undetermined. We evolve $\rho$ using the evolution matrix
$$U =\left(
\matrix{1&~&~&~\cr ~&\rm{cos}\theta&\rm{sin}\theta&~
\cr~&-\rm{sin}\theta&\rm{cos}\theta&~\cr~&~&~&1\cr}\right)$$
and calculate the following constraint equations:
$${\rm tr}_1\left[U\left[
\left(\matrix{\alpha&\beta\cr\beta^*&\gamma\cr}\right) \otimes
\left(\matrix{a&b\cr b^*&c\cr}\right)\right] U^{\dagger}\right] =
\left(\matrix{a&b\cr
b^*&c\cr}\right)~~~~.$$

It turns out that the parameters $a$ and $b$ are given by:$$a = {\left(\alpha -
\alpha
\rm{cos}\theta + 2\mid\beta\mid^2 \rm{cos}\theta\right) \over \left(1 -
\rm{cos}\theta +
4\mid\beta\mid^2 \rm{cos}\theta\right)}~~~~\rm{and}$$ $$b = {\beta
\rm{sin}\theta\left(2a-1\right)\over\left(1-\rm{cos}\theta\right)}$$ {\it i.e.}
there
is a nonlinear
dependence of $a,b$ on $\alpha \rm{~and~} \beta$ so the final state
$\rho_1(T^+)$ will be a nonlinear function of the
initial state and superposition is lost.

\beginsection 6. Conclusions

In this paper, we have examined a proposal for evolution of density matrices
(describing initially pure states) in a spacetime containing closed timelike
curves.
The proposal assumes a tensor product form for the initial quantum state
and a boundary condition imposing consistency of information around the CTC. We
found
that the
evolution of the reduced density matrix describing the state of particles
outside (but
interacting with) the CTC region is described by a nonlinear operator $\Omega$.
This
result poses serious problems for the statistical interpretation of quantum
theory.
In the context of Politzer's example, this is basically the question of what
to take as the initial 4x4 density matrix when the external particle is in a
mixed
quantum state at t=0.

Recall that in the case when the external fermion was in a pure state, the
single tensor
product form was the most general total density matrix such that
${\rm tr}_2\rho = \rho_1$. If $\rho_1$ is mixed, then now the most general
total
density matrix $\rho$ compatible with this condition will not be of this simple
form and it will have a greater number of free parameters. As a consequence,
the boundary condition will not provide enough constraint equations to
determine the final output uniquely.

One could (and Deutsch does) argue that a product form is still justified
because of
the lack of correlation between the 2 spins at t=0. Before t=0, the particles
cannot
interact so why should they be correlated at t=0? However, Politzer noted that
from
the point of view of particle number 2 on the CTC, it has been interacting with
particle 1 in {\it its} past, even if those interactions were not in the other
particle's past. Correlations between the 2 systems need to be taken into
consideration.

So how could we perform the evolution if we did not assume this simple product
form? The standard interpretation of a mixed density matrix $\rho$ is to
decompose it as a statistical ensemble of pure states weighted with
probabilities ie $\rho = \sum c_i\rho_i$, where $\rho_i$ are pure density
matrices and the probabilities $ c_i$ obey $\sum c_i = 1$, $0<c_i<1$. So if
this
interpretation were valid, the evolved mixed state could be unambiguously
determined by summing the results for the evolved pure states which have been
weighted by the initial probabilities $c_i$. This interpretation cannot be used
here
because the initial to final map is nonlinear. Politzer [5] proposed a dynamics
based on a similar decomposition which retains correlations between the
particles at
$t=0$. One of his results is that a pure state evolves to a mixed state even
for
external particles which do not interact with the CTC region. In this example
it is
difficult to see where quantum coherence is being lost as there is no
interaction to
carry the information. If there was a closed timelike curve on the other side
of the
galaxy, it shouldn't affect what happens here. The explanation for this
counterintuitive result is precisely what we have found here - the nonlinear
evolution
ruins any sort of statistical interpretation.

So where does the proposal go wrong? Solving the boundary condition creates the
nonlinear dependence but all the condition really enforces is consistency of
information around the CTC, which seems reasonable. It seems more likely that
the
assumption of a tensor product form as the initial state is where the problem
lies.
The tensor product rules out the possibility of correlations between the 2
systems,
but it seems likely that correlations are indeed present. A more realistic
proposal
may be given based on, for example, the direct sum [7] of Hilbert spaces, in
which the
systems are not assumed independent.

Finally, we consider whether the Deutsch/Politzer time machine is time
symmetric (or
CPT invariant). If strong CPT was obeyed, then one would expect the final total
density matrix $\rho(T^-)$ to be a tensor product when $\rho_2$ is the fixed
point of
\$. However, this is not the case as is readily verified in simple examples.

Weak CPT (or detailed balance) does seem to be obeyed -- that is, the
probability of
transition from A to B is the same as the probability of evolving from the CPT
reverse
of B to the CPT reverse of A. This can be seen in the context of the 2 spin
example
quite easily. Under the CPT operation, a spin up (down) particle becomes a spin
up
(down) antiparticle. The inner product used to calculate transition amplitudes
for
density matrices is essentially just the trace of the dot product of the
matrices so
weak CPT follows from the cyclic property of the trace.

\beginsection Acknowledgements

The author gratefully acknowledges valuable discussions with Chris Fewster and
Stephen
Hawking, and helpful comments from Lloyd Alty, Neil Lambert, Lee London and
Simon Ross.

\beginsection References

\baselineskip=20pt

1. J.L.Friedman, N.J.Papastamatiou and J.Z.Simon, {\it Faliure of Unitarity for
Interacting Fields on Spacetimes with Closed Timelike Curves}, Phys. Rev. {\bf
D46},
 4456 (1992).

2. J.B.Hartle, {\it Unitarity and Causality in Generalised Quantum Mechanics
for
Non-chronal Spacetimes}, gr-qc 9309012.

3. J.B.Hartle, {\it Spacetime Quantum Mechanics and the Quantum Mechanics of
Spacetime} in Proceedings of the 1992 Les Houches Summer School Gravitation and
Quantizations.

4. D.Deutsch, {\it Quantum Mechanics near Closed Timelike Lines}, Phys. Rev.
{\bf D44}
, 3197 (1990).

5. H.D.Politzer, {\it Path Integrals, Density Matrices and Information Flow
with
Closed Timelike Curves}, gr-qc 9310027.

6. A.Anderson, {\it Unitarity in the Presence of Closed Timelike Curves}, gr-qc
9405058.

7. C.J.Fewster and C.G.Wells, {\it Unitarity of Quantum Theory and Closed
Timelike
Curves}, (DAMTP preprint no. R/94/35).

8. S.W.Hawking, {\it The Unpredictability of Quantum Gravity}, Commun. Math.
Phys.
 {\bf 87}, 395 (1982).

\end